# Cross Section Uncertainties in the Gallium Neutrino Source Experiments

W.C. Haxton

*Institute for Nuclear Theory, Box 351550, and Department of Physics, Box 351560*
*University of Washington, Seattle, Washington 98195-1550*
*Email: Haxton@phys.washington.edu*

(October 11, 2018)

## Abstract

The $^{51}$Cr neutrino source experiments play a unique role in testing overall operations of the GALLEX and SAGE solar neutrino experiments. Recently Hata and Haxton argued that the excited-state contribution to the $^{71}$Ga cross section for $^{51}$Cr neutrino absorption might not be known reliably, despite forward-angle (p,n) measurements. A large-basis shell model calculation reported here indicates that the unusual situation they envisioned - destructive interference between weak spin and strong spin-tensor amplitudes - does occur for the transition to the first excited state in $^{71}$Ge. The calculation provides a counterexample to procedures previously used to determine the $^{51}$Cr cross section: the predicted (p,n) cross section for this state agrees with experiment, while the BGT value is well outside the accepted $3\sigma$ limit. The results argue for a shift in the interpretation of the source experiments: they become more crucial as measurements of the $^{71}$Ga detector response to $^7$Be solar neutrinos, and less definitive as wholly independent tests of $^{71}$Ge recovery and counting efficiencies.

Typeset using REVTEX

Recently the GALLEX [1] and SAGE [2] collaborations reported results for test irradiations of their gallium solar neutrino detectors with $^{51}$Cr neutrino sources. Other checks made of detector operations include blank runs; tracer experiments with stable Ge and with Ge carrier doped with $^{71}$Ge; SAGE experiments in which liquid Ga was spiked with the $\beta^-$ sources $^{70}$Ga and $^{72}$Ga, which decay to $^{70}$Ge and $^{72}$Ge; the spiking of the GALLEX detector with the $\beta^+$ source $^{71}$As, which decays to $^{71}$Ge; and the behavior of the detectors during the initial extractions of cosmogenic $^{68}$Ge. Despite these other efforts, the source experiments continue to play a unique role in testing detector operations under conditions where $^{71}$Ge is produced *in situ* and extracted under few-atom, hot chemistry conditions.

The $^{71}$Ge counting rates found in the source experiments depend on the source strength, the overall efficiency for recovering and counting $^{71}$Ge, and the $^{71}$Ga cross section for absorbing $^{51}$Cr neutrinos. As the source activity can be measured to very good accuracy ($\sim 1\%$), the experiment determines the product of the efficiency and cross section. Thus any interpretation of the results as a test of recovery and counting procedures requires strict bounds on cross section uncertainties.

Electron capture on $^{51}$Cr populates two final states in $^{51}$V and thus produces two neutrino lines (neglecting atomic binding energy differences) of energy 746 keV (90%) and 431 keV (10%). An illustrated in Fig. 1, the 431 keV neutrinos can only excite the ground state of $^{71}$Ge, the strength of which is determined by the known 11.43 day lifetime of $^{71}$Ge,

$$\text{BGT(gs)} = \frac{1}{2J_i + 1}|\langle J_f \| O_{\text{GT}}^{J=1} \| J_i \rangle|^2 = 0.087 \pm 0.001 \tag{1}$$

for the Gamow-Teller (GT) matrix element in the direction $^{71}$Ga ($J_i^\pi = 3/2^-$) to $^{71}$Ge ($J_f^\pi = 1/2^-$). The GT operator is

$$O_{\text{GT}}^{J=1} = \sum_{i=1}^{A} \vec{\sigma}(i)\tau_+(i). \tag{2}$$

[One can compare this to Bahcall's recent reanalysis [3] of the $^{71}$Ga decay. Although his results are given in terms of the dimensional cross section factor $\sigma_O = 8.611 \times 10^{-46}$ cm$^2$, one can use Eq. (8.10) of [4], the standard relation

$$ft = \frac{6140 \pm 10}{\text{BF} + g_A^2 \text{BGT}}, \tag{3}$$

and the value $g_A = 1.26$ to derive BGT(gs) = 0.0863, a result consistent with the value in Eq. (1).]

However the dominant 746 keV neutrino branch excites, in addition to the ground state, allowed transitions to the $5/2^-$ and $3/2^-$ states at 175 and 500 keV in $^{71}$Ge. Prior to the paper of Hata and Haxton [5], the excited state transition strengths were thought to be reasonably well known because of forward-angle (p,n) calibrations, which showed that the excited state transitions account for $\sim 5\%$ of the $^{51}$Cr neutrino capture rate. The GALLEX collaboration has used a $^{51}$Cr cross section deduced under this assumption, $5.92 \times 10^{-45}$ cm$^2$, in extracting the ratio R of measured $^{71}$Ge atoms to expected in two source experiments, finding

$$\text{R(GALLEX)} = 1.00 \pm 0.11 \quad \text{and} \quad 0.83 \pm 0.10 \tag{4}$$

The SAGE collaboration has recently quoted its result using a different normalizing $^{51}$Cr cross section of 5.81 ×10$^{-45}$ cm$^2$, yielding

$$\text{R(SAGE)} = 0.95^{+0.11+0.06}_{-0.10-0.05}. \qquad (5)$$

All errors are $1\sigma$.

An alternative normalization of these results is provided by the ground state absorption cross section for $^{51}$Cr neutrinos of $5.53 \times 10^{-45}$ cm$^2$, which can be accurately determined from the $^{71}$Ge lifetime. [This value is taken from Bahcall's recent reevaluation that included a number of improvements, including more accurate atomic wave functions [3].] Combining the two GALLEX measurements and combining the SAGE statistical and systematic errors in quadrature yields

$$\text{R}_0 \equiv \text{E}\left[1 + 0.667\frac{\text{BGT}(5/2^-)}{\text{BGT(gs)}} + 0.218\frac{\text{BGT}(3/2^-)}{\text{BGT(gs)}}\right] = 0.98 \pm 0.08, \text{GALLEX}$$
$$= 1.00 ^{+0.13}_{-0.12}, \text{SAGE} \qquad (6)$$

$\text{R}_0$ is defined as the ratio of the measured counting rate to that expected from the ground state transition only, while the factor E represents any deviation in the overall $^{71}$Ge recovery under source experiment conditions (few-atom, hot-chemistry) from that used by the experimentalists in their analysis. The experimental results on the right-hand side have not been combined because E depends on the experiment: GALLEX and SAGE employ very different chemical procedures. The dependence of $\text{R}_0$ on the unknown transitions strengths BGT($5/2^-$) and BGT($3/2^-$) is explicit and illustrates, in particular, that the $5/2^-$ state will be unimportant only if BGT($5/2^-$) is much smaller than BGT(gs). It is clear at this point that if one wishes to use the source experiment as a test of overall operations of the detector, that is, to check that E = 1, then one must have independent experimental or theoretical arguments constraining the unknown BGT values. The major issue in this paper is to delineate what can be done in this regard, and to point out that the probable situation is quite different from what is conventionally assumed.

Bahcall's recent determination of the $^{51}$Cr cross section was based on the assumption that forward-angle (p,n) measurements provide reliable upper bounds on weak BGT values. I show below that this is not generally true. Furthermore a sophisticated shell model calculation is performed which demonstrates that this is not the case for the transition to the first excited state in $^{71}$Ge. The calculation predicts destructive interference between the (p,n) spin and spin-tensor matrix elements, the possibility envisioned in Hata and Haxton [5]. I discuss how this result affects the interpretation of the results of the source experiments.

The Krofcheck et al. [6] (p,n) measurements for $^{71}$Ga were made at 120 and 200 MeV, yielding

$$\text{BGT}^{\text{exp}}_{(\text{p,n})}(5/2^-) < 0.005 \text{ and } \text{BGT}^{\text{exp}}_{(\text{p,n})}(3/2^-) = 0.011 \pm 0.002. \qquad (7)$$

From these results the 5% estimate of excited state contributions to the source experiments was deduced. However, while the reliability of forward-angle (p,n) reactions in mapping the overall BGT strength profile of nuclei is reasonably well established, discrepancies in the case of individual transitions of known strength have been noted. Table I, repeated from Ref. [5], compares 10 transitions for which both $\beta$ decay and (p,n) information is available. In over half of these cases, the $\beta$ decay and (p,n) BGT values disagree significantly.

As discussed in Refs. [7,8], the underlying reason for the discrepancies in Table I appears to be the presence of a spin-tensor (L=2 S=1)J=1 component in the forward-angle (p,n) operator,

$$\langle J_f \| O_{(p,n)}^{J=1} \| J_i \rangle = \langle J_f \| O_{GT}^{J=1} \| J_i \rangle + \delta \langle J_f \| O_{L=2}^{J=1} \| J_i \rangle_{SM} \tag{8}$$

where

$$O_{L=2}^{J=1} = \sqrt{8\pi} \sum_{i=1}^{A} [Y_2(\Omega_i) \otimes \vec{\sigma}(i)]_{J=1} \tau_+(i). \tag{9}$$

and where the notation $\langle \| \ \| \rangle_{SM}$ indicates that a shell model reduced matrix element is to be taken. Thus $BGT_{(p,n)}^{SM}$ is defined in analogy with Eq. (1), but with the operator in Eq. (8) replacing that in Eq. (2). This effective operator indeed proves to remove all of the large discrepancies in Table I, provided one takes $\delta \sim 0.85$. (The fitted values used in the table are 0.069 and 0.096 for the 2s1d and 1p shells, respectively.) The resulting values $BGT_{(p,n)}^{SM}$ one then obtains, listed in the last column of Table I, are in good agreement with the measured (p,n) values. [The calculations were done by using the $\beta$ decay value for the magnitude of $\langle J_f \| O_{GT}^{J=1} \| J_i \rangle$ and shell model values for $\langle J_f \| O_{L=2}^{J=1} \| J_i \rangle$ and for the relative sign of the matrix elements [5].]

This discussion shows that the tendancy in Table I for (p,n) reactions to overestimate true BGT values does not reflect some general property of (p,n) reactions, but rather a specific common property of these transitions: the transition densities are dominantly diagonal, either of the form $|(l\frac{1}{2})j = l - \frac{1}{2}\rangle \to |(l\frac{1}{2})j = l - \frac{1}{2}\rangle$ or $|(l\frac{1}{2})j = l + \frac{1}{2}\rangle \to |(l\frac{1}{2})j = l + \frac{1}{2}\rangle$. In Table II we show that, for mirror transitions and others of this character, the interference between the GT and spin-tensor operators in Eq. (8) is constructive. But Table II also shows that even for pure single-particle transitions, destructive interference can result, as in the case of transitions between spin aligned and spin antialigned configurations. This will generally result in a (p,n) BGT value that is smaller than the true $\beta$ decay BGT value. Furthermore, below we will explicitly show that a (p,n) BGT value can be *substantially* smaller than the true value - with the transition to the $5/2^-$ state in $^{71}$Ge being a very likely example.

Hata and Haxton [5] pointed out that simple descriptions of $^{71}$Ga and $^{71}$Ge - a Nilsson model with modest positive deformation $\beta \sim 0.05 - 0.15$ accounts for $1/2^-, 5/2^-, 3/2^-$ level ordering in $^{71}$Ge - suggest that the density matrix for the $3/2^- \to 5/2^-$ transition to the first excited state in $^{71}$Ge is likely dominated by

$$1f_{5/2}(n) \to 2p_{3/2}(p). \tag{10}$$

This is an *l*-forbidden M1 transition, an example of the fourth category in Table II, and similar to the $^{39}$K $\to$ $^{39}$Ca$(1/2^+)$ case in Table I. This particular transition generates the largest spin-tensor matrix element in the 2p1f shell: if the transition were of single-particle strength, the resulting $BGT_{(p,n)}$ would be an order of magnitude larger than the experimental upper bound. This could indicate that the $1f_{5/2} \to 2p_{3/2}$ amplitude, unlike the simple one-hole $^{39}$K case, is considerably below single-particle strength. But a second possibility, for a more complex transition of this sort involving nuclei in the middle of a shell, is that the small $BGT_{(p,n)}$ comes about through a cancellation between the GT and spin-tensor operators. The competing GT amplitude would arise from presumably less important terms in the density matrix, e.g., $2p_{1/2} \to 2p_{3/2}$ and $1f_{5/2} \to 1f_{5/2}$. If this were the case, the $\beta$ decay BGT value could be considerably larger than the (p,n) bound. When Hata and Haxton explored this issue in detail, they found values of $BGT(5/2^-)$ between 0 and BGT(gs) could still be compatible with the (p,n) constraint, given conceivable values for the strength of the unknown spin-tensor matrix element. Thus $R_0$ is only weakly constrained to the range 1 to 1.667, a cross section uncertainty that would make the source experiments much less useful as a test of detector operations.

Hata and Haxton limited their investigations to delineating what might be possible: no effort was made to use nuclear theory to try to limit these possibilities, i.e., to determine what might be probable. The discussion of the relationship between BGT and $BGT_{(p,n)}$ in the 1p and 2s1d shells should then be encouraging. If one makes no use of theory in Table I, large discrepancies appear between $\beta$ decay and (p,n) BGT determinations. But the inclusion of the spin-tensor operator, which theory tells us should be present in the (p,n) amplitude, combined with standard shell model evaluations of the relative sign and magnitude of this second operator, nicely removes all large discrepancies between $\beta$ decay and (p,n) BGT evaluations. Below we follow the same strategy in the case of $^{71}$Ga.

However this involves a complication as unconstrained shell model calculations in the canonical shell model space $(2p_{3/2}1f_{5/2}2p_{1/2}1g_{9/2})$ for $^{71}$Ga and $^{71}$Ge are still somewhat out of reach, unlike the 1p and 2s1d shell cases of Table I. The necessary truncations of this space cannot be too violent due to the deformation effects apparent in this mass region. For example, the energy of the first excited $0^+$ state in the lighter even-A isotopes of Ge plunges as the number of neutrons is increased, apparently leading to a level crossing with the ground state at neutron number $\sim 40$. The proton occupation numbers, derived from measured spectroscopic factors, are changing rapidly at the same point. The $2p_{3/2}$ occupation drops dramatically as the $1f_{5/2}$ occupation rises. As discussed in Ref. [9], these rather dramatic structure changes are reproduced by a weak-coupling shell model, from which the underlying physics can be extracted. As neutrons begin to occupy the $1g_{9/2}$ shell, a strong polarizing interaction arises between $1g_{9/2}$ neutrons and $1f_{5/2}$ protons: these orbits have the same nodal structure and thus have favorable spatial overlap. The interaction has a strong influence on the structure of the ground state as one approaches the naive N=40 closed neutron shell. An examination of the largest wave function components in the calculation of Ref. [9] shows that the spherical proton configuration $2p_{3/2}^4$ is admixed with the deformed configurations $2p_{3/2}^2 1f_{5/2}^2$ and $2p_{3/2}^0 1f_{5/2}^4$, while the spherical neutron configuration $2p_{1/2}^2$ becomes admixed with the deformed configuration $2p_{1/2}^0 1g_{9/2}^2$. The transition from an essentially spherical ground state at N=38 to a deformed ground state at N=40 is particularly sharp because it is driven by the strong $1f_{5/2}(p)$-$1g_{9/2}(n)$ attraction, which favors the deformation, and leads to premature occupation of the $1g_{9/2}$ shell. This interpretation is consistent with the Nilsson model, where an orbital associated with the spherical $1g_{9/2}$ shell plunges below the Nilsson orbital associated with the $2p_{1/2}$ shell for large positive deformation.

It is clear at this point that a realistic shell model calculation of the N=40 nucleus $^{71}$Ga must include the excitations into the $1g_{9/2}$ shell that drive deformation. Such a calculation is now practical: the inclusion of all configurations of the form $(2p_{3/2}1f_{5/2}2p_{1/2})^{15}1g_{9/2}^0$ and $(2p_{3/2}1f_{5/2}2p_{1/2})^{13}1g_{9/2}^2$ results in a m-scheme basis for $^{71}$Ge of about 492,000. Matrices of this dimension can be handled with relative ease on a large-memory workstation. The calculation was performed using the interaction of Ref. [10], with single-particle energies adjusted to fit the level ordering in $^{71}$Ge and $^{71}$Ga.

The resulting shell model matrix elements and predicted $\beta$ decay and (p,n) BGT values are given in Table III. The one-body density matrix for the transition to the $1/2^-$ ground state of $^{71}$Ge is dominated by the $2p_{1/2}(n) \to 2p_{3/2}(p)$ amplitude, and thus corresponds to the third possibility in Table II (and is distinct from any of the cases in Table I). Consequently the GT and spin-tensor operators are predicted to interfere destructively, leading to a $BGT_{(p,n)}$ that is slightly smaller than the corresponding $\beta$ decay value. (We use $\delta = 0.097$ in the 2p1f shell [8].) The predicted $BGT_\beta$ of 0.051 is in quite reasonable agreement with experiment (0.087), corresponding to shell model matrix element of $\sim 0.77$ the experimental value.

Although the experimental and calculated (p,n) BGT values for the $3/2^-$ disagree numerically, both values are small, 0.011 and 0.0011, respectively.

But the remarkable entry in Table III is that for the transition to the $5/2^-$ first excited state. The transition density is dominated by the $l$-forbidden $1f_{5/2}(n) \to 2p_{3/2}(p)$ amplitude, leading to a huge spin-tensor operator matrix element. (The calculated value corresponds to 0.48 of the single-particle value.) The next most important contribution to the transition density, $2p_{1/2}(n) \to 2p_{1/2}(p)$, generates a small GT matrix element that interferes destructively with the spin-tensor matrix element. Note that the final (p,n) BGT value, 0.0006, is in agreement with the experimental upper bound of 0.005.

Now the use of these results depends on one's goals. I feel there are three logical ways of proceeding:

*i) Testing the overall operations of the GALLEX and SAGE detectors.* If the goal is to use the experimental constraints in Eq. (6) to derive a bound on E, clearly an independent constraint is needed on the excited state BGT values. The standard procedure has been to employ the experimental (p,n) BGT values (Eq.(7)) in Eq. (6), which yields the result

$$\begin{aligned} E &= 0.94 \pm 0.08 \pm 0.02, \quad \text{GALLEX} \\ &= 0.96 \, ^{+0.13}_{-0.12} \pm 0.02, \quad \text{SAGE} \end{aligned} \quad (11)$$

where the second uncertainty reflects the experimental uncertainty in the measured (p,n) BGT values (Eq. (7)). But this procedure - equating the $\beta$ decay BGTs to the (p,n) values - is *clearly* not defensible: the nuclear structure study described above predicts a $5/2^-$ (p,n) BGT value in agreement with experiment, but yields a $\beta$ decay BGT value almost four times larger than would be allowed in this simplistic analysis.

The approach taken in Hata and Haxton was to allow the GT and spin-tensor matrix elements to take on any values consistent with the (p,n) results and the constraint that the strength of the spin-tensor matrix element could not exceed the single-particle limit. Now that we have a reasonable theoretical description of the $^{71}$Ga weak and (p,n) transitions, we have some chance to narrow this range. Because the spin-tensor transition to the $5/2^-$ state is so strong, the obvious strategy is to mimic the calculations summarized in Table I: use theory to predict the magnitude and relative sign of the spin-tensor amplitude, then limit the GT amplitude by using Eqs. (7-9). This is clearly preferable to directly calculating the GT matrix element, which the shell model predicts is almost a factor of 10 smaller than the spin-tensor matrix element. The net result is

$$0.0014 < \text{BGT}(5/2^-) < 0.032. \tag{12}$$

For the $3/2^-$ state it is reasonable to adopt the (p,n) BGT value, as the shell model predicts this is a typical transition where the (p,n) and $\beta$ decay values are comparable. A short calculation then yields

$$\begin{aligned} E &= 0.86 \pm 0.07 \pm 0.09, \quad \text{GALLEX} \\ &= 0.875 \pm 0.11 \pm 0.09, \quad \text{SAGE}, \end{aligned} \tag{13}$$

where the second error represents the BGT uncertainty of Eq. (12). Note that if the directly calculated shell model BGT($5/2^-$) is used, 0.017, the resulting Es are in the middle of these ranges, 0.85 and 0.86, respectively. [This shell model value is in good agreement with the earlier estimate by Mathews et al. [11] (0.020), even though this calculation did not include the important deformation effects associated with the $1g_{9/2}$ shell. This may not be accidental: among the $\sim 20$ low-lying states in $^{71}$Ga and $^{71}$Ge that converged in our shell model study, the $^{71}$Ga ground state and the $5/2^-$ $^{71}$Ge first excited state had the smallest occupation of the $1g_{9/2}$ shell.]

I regard Eq. (13) as the best current statement about the implications of the source experiments for the overall operations of GALLEX and SAGE. The ranges include E $\sim$ 1: there is no indication of any operational problem. But substantial variations from E $\sim$ 1 are also allowed. One of the features of Eq. (13) is that the theory error is comparable to the precision of the experiments. Thus further improvements in the source experiments will not tighten the constraints on E unless some progress is made on the excited state nuclear structure uncertainties.

*ii) Reducing errors in derived solar neutrino fluxes.* The $^{71}$Ga detector response to various neutrino sources depends on quantities such as

$$\text{E}\langle\sigma\phi(\text{pp})\rangle \quad \text{E}\langle\sigma\phi(^7\text{Be})\rangle \quad \text{E}\langle\sigma\phi(^8\text{B})\rangle. \tag{14}$$

The pp cross section is almost entirely due to the ground state transition. In the case of $^8$B neutrinos, the cross section is quite uncertain, with the best determination coming from the (p,n) mapping of the bound-state BGT profile in $^{71}$Ge [6]. But Eq. (12) then limits the contributions of the 175 and 500 keV states to less than 6% of the total cross section [3]. Thus the first two excited states do not contribute appreciably to estimated uncertainties in the pp and $^8$B neutrino gallium responses. Of course, the extraction of E, discussed above, is important to these predictions.

But the $^7$Be response is governed by the same transitions that are involved in the $^{51}$Cr source experiment. Eq. (15) of Hata and Haxton can be rewritten as

$$\begin{aligned} \langle\sigma\phi(^7\text{Be})\rangle = \text{E}(1.3\text{SNU})&\text{P}_{\text{MSW}}(384\text{keV}) + \\ \text{R}_0(34.4\text{SNU})&\text{P}_{\text{MSW}}(862\text{keV})\frac{\text{BGT(gs)} + 0.711\text{BGT}(5/2^-) + 0.290\text{BGT}(3/2^-)}{\text{BGT(gs)} + 0.667\text{BGT}(5/2^-) + 0.218\text{BGT}(3/2^-)} \end{aligned} \tag{15}$$

where the possibility of neutrino oscillations is included through the factors $P_{MSW}$, which give the ratio of the flux with oscillations to that without for the two $^7$Be lines at 384 and 863 keV. A $^7$Be flux of 5.15E9/cm$^2$s has been used, corresponding to the Bahcall and Pinsonneault standard solar model with He and metal diffusion [12]. The strong similarities between the $^{51}$Cr and $^7$Be neutrino spectra were exploited to replace the E and the unknown nuclear structure quantities by a *measured* quantity $R_0$, leaving a residual nuclear structure factor

$$\frac{\text{BGT(gs)} + 0.711\text{BGT}(5/2^-) + 0.290\text{BGT}(3/2^-)}{\text{BGT(gs)} + 0.667\text{BGT}(5/2^- + 0.218\text{BGT}(3/2^-)} = 1.012 \pm 0.004 \tag{16}$$

which proves to be remarkably constant when $\text{BGT}(5/2^-)$ and $\text{BGT}(3/2^-)$ are allowed to vary over the full ranges given by Eqs. (12) and (7), respectively.

Thus Eq. (15) allows one to predict the GALLEX and SAGE responses to a given flux of $^7$Be neutrinos, almost independent of uncertainties in E or in excite state BGT values, given accurate measurements of $R_0$. Unlike our conclusion in *i)*, this relation provides strong motivation for further source experiments to reduce the error in $R_0$.

*iii) The $^{51}$Cr cross section.* In this section we gather together various determinations, with cautionary comments, of the excited state BGT values or, almost equivalently, the $^{51}$Cr cross section.

If one is willing to stipulate that E $\sim$ 1, the GALLEX and SAGE experiments then require (see [3,5])

$$0.667\frac{\text{BGT}(5/2^-)}{\text{BGT(gs)}} + 0.218\frac{\text{BGT}(3/2^-)}{\text{BGT(gs)}} = -0.02 \pm 0.08, \quad \text{GALLEX}$$
$$= 0.00^{+0.13}_{-0.12}, \quad \text{SAGE} \tag{17}$$

This result is helpful, as in *i)*, in showing that the source experiments and the assumption E $\sim$ 1 are compatible with a reasonable range of excited state BGT values. However, it does not provide a useful basis for deriving a $^{51}$Cr cross section, as the subsequent use of this cross section in analyzing the source experiments would then be a tautology.

To be relevant to the source experiment, the cross section must be derived from information independent of that experiment. Thus the (p,n) results must be used and, as we showed in Table I and especially in the case of the $5/2^-$ state in $^{71}$Ge, the relationship between (p,n) cross sections and the corresponding BGTs must take into account the complicating effects of the spin-tensor operator. The procedures used in Table I can fortunately be extended to $^{71}$Ge because the spin-tensor matrix element is predicted to be so strong, and thus hopefully can be calculated with a degree of success similar to the cases in the Table. Thus using Eq. (12) and, as argued previously, the second of Eqs. (7), one finds

$$0.667\frac{\text{BGT}(5/2^-)}{\text{BGT(gs)}} + 0.218\frac{\text{BGT}(3/2^-)}{\text{BGT(gs)}} = 0.15 \pm 0.12 \tag{18}$$

yielding [13]

$$\sigma(^{51}\text{Cr}) = (6.39 \pm 0.68) \times 10^{-45}\text{cm}^2. \tag{19}$$

This can be compared to the corresponding result where $\text{BGT}(5/2^-)$ is taken directly from our shell model calculation

$$\sigma(^{51}\text{Cr}) = 6.41 \times 10^{-45}\text{cm}^2 \tag{20}$$

and to the recent result of [3]

$$\sigma(^{51}\text{Cr}) = (5.81^{+0.21}_{-0.16}) \times 10^{-45} \text{cm}^2. \tag{21}$$

The error in Eq. (21) includes uncertainties from forbidden corrections and from the $^{71}$Ga threshold and lifetime. The portion of the error associated with excited state uncertainties, appropriate for comparison with Eq. (19), is $^{+0.16}_{-0.09}$.

The very narrow range in Eq. (21) results from arguments that (p,n) BGT values should be upper bounds to the true weak interaction values, based on the trends in Table I. Unfortunately we have seen that constructive interference between the GT and spin-tensor operators is not a general feature of (p,n) reactions, but rather of diagonal transition densities, such as occur for the mirror or nearly mirror transitions that dominate Table I. The shell model result reported here provides an explicit counterexample in the case of most interest to us, the $5/2^-$ state. This calculation predicts a BGT($5/2^-$) that is far outside the $3\sigma$ range considered in [3], yet is in agreement with the (p,n) value, the same input used in [3]. The resulting $\sigma(\text{Cr}^{51})$ (Eq. (20)) is $\sim 3\sigma$ from the value of Eq. (21). The range in Eq. (19) extends to $\sim 6\sigma$. Finally, it could be argued that the range in Eq. (19) is still too conservative, as it does not taken into account theoretical uncertainties in the evaluation of the spin-tensor matrix element or in the value adopted for $\delta$, which are very difficult to quantify.

The results presented in this paper provide motivation for more careful experimental studies of the (p,n) cross section for the $5/2^-$ state. The (p,n) energy and angular dependence and new spin-transfer measurement could help to separate the spin and spin-tensor contributions. One existing measurement provides some support for the shell model description presented here. The anomalously strong $5/2^-$ (p,n) cross section found at 35 MeV, comparable to the ground state cross section, was attributed to a strong spin-tensor contribution [14]: the spin-tensor contribution is expected to increase in importance as the proton energy decreases.

I thank Eric Adelberger, John Bahcall, Tony Baltz, Steve Elliott, Virginia Brown, Dick Hahn, and John Wilkerson for helpful discussions. This work was supported in part by the US Department of Energy.

TABLES

TABLE I. Comparison of $\beta$ decay BGT values, experimental (p,n) BGT values, and $\text{BGT}^{\text{SM}}_{(p,n)}$ calculated from the effective operator of Eq. (8), using $\delta = 0.069$ (0.096) for the 2s1d (1p) shell. See Ref. [5] for additional information.

| Nucleus | $J_i$ | $J_f(E_f(\text{MeV}))$ | $\text{BGT}^{exp}_{\beta}$ | $\text{BGT}^{exp}_{(p,n)}$ | $\text{BGT}^{\text{SM}}_{(p,n)}$ |
|---|---|---|---|---|---|
| $^{13}$C | $1/2^-$ | $1/2^-$ (0.0) | 0.20 | 0.39 | 0.40 |
| $^{14}$C | $0^+$ | $1^+$ (3.95) | 2.81 | 2.82 | 2.84 |
| $^{15}$N | $1/2^-$ | $1/2^-$ (0.0) | 0.25 | 0.54 | 0.53 |
| $^{17}$O | $5/2^+$ | $5/2^+$ (0.0) | 1.05 | 0.99 | 1.15 |
| $^{18}$O | $0^+$ | $1^+$ (0.0) | 3.06 | 3.54 | 3.11 |
| $^{19}$F | $1/2^+$ | $1/2^+$ (0.0) | 1.62 | 2.13 | 1.65 |
| $^{26}$Mg | $0^+$ | $1^+$ (1.06) | 1.10 | 1.14 | 1.20 |
| $^{32}$S | $0^+$ | $1^+$ (0.0) | 0.0021 | 0.014 | 0.016 |
| $^{39}$K | $3/2^+$ | $3/2^+$ (0.0) | 0.27 | 0.39 | 0.39 |
| $^{39}$K | $3/2^+$ | $1/2^+$ (2.47) | 0.00017 | $\sim$0.017 | 0.014 |

TABLE II. The matrix element ratio $\langle f\|O^{J=1}_{L=2}\|i\rangle/\langle f\|O^{J=1}_{\text{GT}}\|\rangle$ for single-particle transitions. The last column classifies the transitions in Table I according to their dominant character.

| $\|f\rangle$ | $\|i\rangle$ | Ratio | Examples |
|---|---|---|---|
| $\|(l\frac{1}{2})j=l-\frac{1}{2}\rangle$ | $\|(l\frac{1}{2})j=l-\frac{1}{2}\rangle$ | 2(l+1)/(2l-1) | $^{13}$C($1p_{1/2} \to 1p_{1/2}$) |
| | | | $^{14}$C($1p_{1/2} \to 1p_{1/2}$) |
| | | | $^{15}$N($1p_{1/2} \to 1p_{1/2}$) |
| | | | $^{39}$K($1d_{3/2} \to 1d_{3/2}$) (0.0 MeV) |
| $\|(l\frac{1}{2})j=l+\frac{1}{2}\rangle$ | $\|(l\frac{1}{2})j=l+\frac{1}{2}\rangle$ | 2l/(2l+3) | $^{17}$O($1d_{5/2} \to 1d_{5/2}$) |
| | | | $^{18}$O($1d_{5/2} \to 1d_{5/2}$) |
| | | | $^{19}$F($2s_{1/2} \to 2s_{1/2}$) |
| | | | $^{26}$Mg($1d_{5/2} \to 1d_{5/2}$) |
| | | | $^{32}$S($1d_{5/2} \to 1d_{5/2}$) |
| $\|(l\frac{1}{2})j=l-\frac{1}{2}\rangle$ | $\|(l\frac{1}{2})j=l+\frac{1}{2}\rangle$ | -1/2 | |
| $\|(1\frac{1}{2})j=l+\frac{1}{2}\rangle$ | $\|((l+2)\frac{1}{2})j=l+\frac{3}{2}\rangle$ | $\pm\infty$ | $^{39}$K($1d_{3/2} \to 2s_{1/2}$) (2.47 MeV) |

TABLE III. Large-basis shell model results for $^{71}$Ga $\to$ $^{71}$Ge Gamow-Teller and spin-tensor matrix elements and the corresponding BGT predictions. The (p,n) BGT calculation was done for $\delta = 0.097$.

| Transition | $\langle f\|O_{GT}\|i\rangle$ | $\langle f\|O_{L=2}\|i\rangle$ | BGT$_\beta^{\text{SM}}$ | BGT$_{(p,n)}^{\text{SM}}$ |
|---|---|---|---|---|
| $3/2^- \to 1/2^-$ (0 keV) | -0.451 | 0.348 | 0.051 | 0.044 |
| $3/2^- \to 5/2^-$ (175 keV) | 0.264 | -2.23 | 0.017 | 0.0006 |
| $3/2^- \to 3/2^-$ (500 keV) | 0.056 | 0.104 | 0.0008 | 0.0011 |

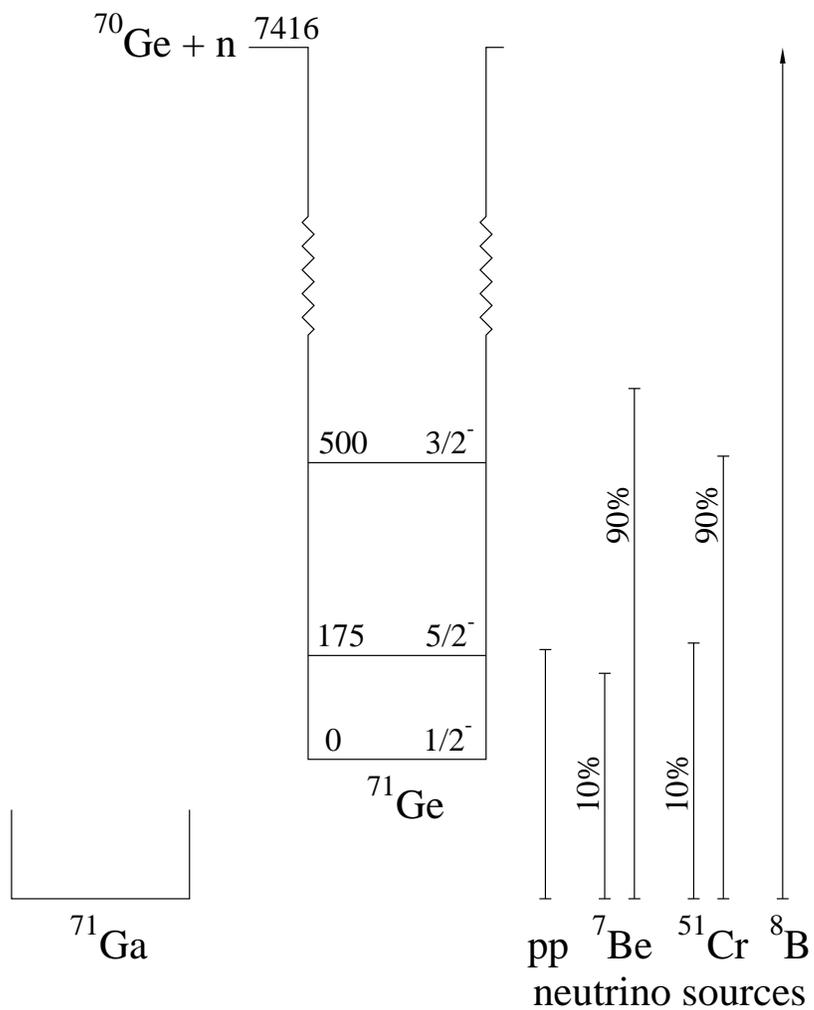

FIG. 1. Level scheme for $^{71}$Ge showing the excited states that contribute to absorption of pp, $^7$Be, $^{51}$Cr, and $^8$B neutrinos.